\documentclass[a4paper]{jpconf}

\usepackage{graphicx}
\usepackage[english]{babel}
\usepackage{amsmath,amssymb}

\def \1{\hbox{\small 1\kern-3.6pt\normalsize 1}}

\begin{document}
\title{Sterile neutrinos and $R_K$}

\author{A.~Vicente}

\address{Laboratoire de Physique Th\'eorique, CNRS -- UMR 8627,
Universit\'e de Paris-Sud 11, F-91405 Orsay Cedex, France}

\ead{avelino.vicente@th.u-psud.fr}

\begin{abstract}
We study the violation of lepton flavour universality in light meson
decays due to the presence of non-zero mixings between the active
neutrinos with new sterile states. The modified $W \ell \nu$ vertices,
arising from a non-unitarity leptonic mixing matrix intervening in
charged currents, might lead to a tree-level enhancement of $R_{P} =
\Gamma (P \to e \nu) / \Gamma (P \to \mu \nu)$, with $P=K, \pi$. These
enhancements are illustrated in the case of the inverse seesaw,
showing that one can saturate the current experimental bounds on
$\Delta r_{K}$ (and $\Delta r_{\pi}$), while in agreement with the
different experimental and observational constraints.
\end{abstract}

\section{Introduction}\label{Sec:Introduction}
Lepton flavour universality (LFU) is a distinctive feature of the
Standard Model (SM). The different lepton families couple with exactly
the same strength to the gauge bosons. This leads to concrete
predictions in electroweak precision tests, which can only distinguish
among lepton families by the different charged lepton masses. Any
deviation from the expected SM theoretical results would signal the
presence of New Physics (NP).  In this work we concentrate on light
meson ($K$ and $\pi$) leptonic decays which, in view of the expected
experimental precision, have a unique potential to probe deviations
from the SM regarding lepton universality.

In the SM, the dominant contribution to $\Gamma (P \to \ell \nu)$
($P=K, \pi$) arises from $W$ boson exchange. One may be
afraid about potential hadronic uncertainties; however, by considering
the ratios
\begin{equation}\label{eq:RK:Rpi}
R_K \, \equiv \,\frac{\Gamma (K^+ \to e^+ \nu)}{\Gamma (K^+ \to \mu^+
  \nu)}\,,
\quad \quad
R_\pi \, \equiv \,\frac{\Gamma (\pi^+ \to e^+ \nu)}{\Gamma (\pi^+ \to \mu^+
  \nu)}\,, 
\end{equation}
the hadronic uncertainties are expected to cancel out to a good
approximation.  In order to compare the experimental bounds with the
SM predictions, it is convenient to introduce a quantity, $\Delta
r_P$, which parametrizes deviations from the SM expectations:
\begin{equation}\label{eq:deltar:P}
R_P^\text{exp} \, = \,R_P^\text{SM} \, (1+\Delta r_P) \quad
\text{or equivalently}
\quad
\Delta r_P \, \equiv \, \frac{R_P^\text{exp}}{R_P^\text{SM}} - 1\,.
\end{equation}
The comparison of theoretical
analyses~\cite{Cirigliano:2007xi,Finkemeier:1995gi} with the recent
measurements from the NA62 collaboration~\cite{Goudzovski:2011tc} and
with the existing measurements on pion leptonic
decays~\cite{Czapek:1993kc}
\begin{eqnarray}\label{eq:RK:Rpi:SMvsexp}
& R_K^\text{SM} \, = \, (2.477 \pm 0.001) \, \times 10^{-5}\,,
\quad \quad
& R_K^\text{exp} \, = \,(2.488 \pm 0.010) \, \times 10^{-5}\,,
\\
& R_\pi^\text{SM} \, =\, (1.2354 \pm 0.0002) \, \times 10^{-4}\,, 
\quad \quad
& R_\pi^\text{exp} \, =\, (1.230 \pm 0.004) \, \times 10^{-4}\,
\end{eqnarray}
suggests that observation agrees at $1 \sigma$ level with 
the SM's predictions for 
\begin{equation}\label{eq:deltar:P:value} 
\Delta r_K \, = \, (4 \pm 4 )\, \times\, 10^{-3}\,, 
\quad \quad
 \Delta r_\pi \, = \, (-4 \pm 3 )\, \times\, 10^{-3}\,. 
\end{equation}
The current experimental uncertainty in $\Delta r_K$ (of around 0.4\%)
will be further reduced in the near future, as one expects to have
$\delta R_K / R_K \sim 0.1\%$~\cite{Goudzovski:2012gh}, which can
translate into measuring deviations $\Delta r_K \, \sim
\mathcal{O}(10^{-3})$. Similarly, there are also plans for a more precise
determination of $\Delta
r_\pi$~\cite{Pocanic:2012gt,Malbrunot:2012zz}.

New contributions to $\Delta r_P$ have been extensively discussed in
the literature, especially in the framework of models with an enlarged
Higgs sector. In the presence of charged scalar Higgs, new tree-level
contributions are expected. However, as in the case of most Two
Higgs Doublet Models (2HDM), or supersymmetric (SUSY) extensions of
the SM, these new tree-level corrections are lepton
universal~\cite{Hou:1992sy}.  In SUSY models, higher order
non-holomorphic couplings can indeed provide new contributions to
$R_P$~\cite{Masiero:2005wr,Masiero:2008cb,Ellis:2008st,Girrbach:2012km,Fonseca:2012kr},
but in view of current experimental bounds (collider, $B$-physics and
$\tau$-lepton decays), one can have at most $\Delta r_K \leq 10^{-3}$
in the framework of unconstrained minimal SUSY
models~\cite{Fonseca:2012kr}. Corrections to the $W \ell \nu$ vertex
can also induce violation of LFU in charged currents. However, if
these appear at the loop level, as referred to
in~\cite{Masiero:2008cb}, the effect is expected to be of order
$(\alpha/4 \pi) \times (m^2_W/\Lambda^2_\text{NP})$
($\Lambda_\text{NP}$ being the new physics scale), generally well
below experimental sensitivity.

Here we consider a different alternative: The tree-level corrections
to charged current interactions once neutrino oscillations are
incorporated into the SM \cite{Shrock,Abada:2012mc}.  In this case, and
working in the basis where the charged lepton mass matrix is diagonal,
the flavour-conserving term $\propto g \bar{l}_j \gamma^\mu P_L \nu_j
W_\mu^-$ now reads
\begin{equation}\label{eq:cc-lag}
- \mathcal{L}_{cc} = \frac{g}{\sqrt{2}} U_\nu^{ji} 
\bar{l}_j \gamma^\mu P_L \nu_i  W_\mu^- + \, \text{c.c.}\,,
\end{equation}
where $U_\nu^{ji}$ is a generic leptonic mixing matrix, $i = 1, \dots,
n_\nu$ denoting the physical neutrino states (not necessarily
corresponding to the three left-handed SM states $\equiv \nu_L$) and
$j = 1, \dots, 3$ the charged lepton flavour.  In the case of three
neutrino generations, $U_\nu^{ji}$ corresponds to the unitary PMNS
matrix and flavour universality is preserved in meson decays: since
one cannot tag the flavour of the final state neutrino (missing
energy), the meson decay amplitude is proportional to $(U_{\nu}
U_\nu^\dagger)_{jj} =1$, and thus no new contribution to $R_P$ is
expected. However, in the presence of sterile states, the $W \ell \nu$
vertex is proportional to a rectangular $ 3\times n_\nu$ matrix
$U_\nu^{ji}$, and the mixing between the left-handed leptons $\nu_L,
\ell_L$ corresponds to a $3 \times 3$ block of $U_\nu^{ji}$,
\begin{equation}\label{eq:U:eta:PMNS}
U_\text{PMNS} \, \to \, \tilde U_\text{PMNS} \, = \,(\1 - \eta)\, 
U_\text{PMNS}\,.
\end{equation}
The larger the mixing between the active (left-handed) neutrinos and
the new states, the more pronounced the deviations from unitarity of
$\tilde U_\text{PMNS}$, parametrized by the matrix
$\eta$~\cite{eta}. The active-sterile mixings and the departure from
unitarity of $\tilde U_\text{PMNS}$ can be at the source of the
violation of LFU in different neutrino mass models which introduce
sterile fermionic states to generate non-zero masses and mixings for
the light neutrinos \cite{Shrock,Abada:2012mc}.

Corrections to the $W \ell \nu$ vertex can arise in several scenarios
with additional (light) singlet states, as is the case of
$\nu$SM~\cite{Asaka:2005an}, low-scale type-I
seesaw~\cite{Ibarra:2010xw} and the Inverse Seesaw
(ISS)~\cite{Mohapatra:1986bd}, among other possibilities. This clearly
shows the potentiality of the mechanism under discussion, which can be
present in many different models.

In the next section we provide a model-independent computation of
$\Delta r_{P}$ in the presence of additional fermionic sterile states;
we then briefly review in Section~\ref{Sec:Constraints} the most
important experimental and observational constraints on the mass of
the additional singlet states.  In Section~\ref{Sec:drkiss}, we
consider the case of the inverse seesaw to give a numerical
example of the impact of sterile neutrinos on $\Delta r_{P}$. Our
concluding remarks are summarised in Section~\ref{Sec:conclusions}.

\section{$\Delta r_P$ in the presence of sterile neutrinos}
\label{Sec:drk}

Let us consider the SM extended by $N_s$ additional sterile states.
The matrix element for the meson decay $P \to l_j \nu_i$ can be
generically written as
\begin{equation} \label{eq:mat-elem}
\mathcal{M}_{ij} = \bar{u}_{\nu_i} (\mathcal{A}^{ij} P_R + 
\mathcal{B}^{ij} P_L ) v_{l_j} \,.
\end{equation}
No sum is implied over the indices of the outgoing leptons
$i,j$. Notice that $i=1, \dots, 3+N_s$. The expressions for
$\mathcal{A}$ and $\mathcal{B}$ can be easily obtained from the usual
4-fermion effective hamiltonian obtained after integrating out the $W$
boson in Eq. \eqref{eq:cc-lag}. These are
\begin{align}\label{eq:eff-ham2}
& (\mathcal{A})^{ij} \, =\, (\mathcal{A}^W)^{ij}\, =\, 
- 4 \,G_F \,V_\text{CKM}^{us} \, f_P \,U_\nu^{ji \, *}\,  m_{l_j} \,;\\
& (\mathcal{B})^{ij} \, =\, (\mathcal{B}^W)^{ij}\, =\, 
4 \,G_F \,V_\text{CKM}^{us} \,f_P \,U_\nu^{ji \, *}\,  m_{\nu_i}\,,
\end{align}
where $f_P$ denotes the meson decay constant and $m_{l_j, \nu_i}$ the
mass of the outgoing leptons.

The expression for $R_P$ is finally given by
\begin{equation}\label{eq:RPresult}
R_P \,= \,\frac{\sum_i 
F^{i1} G^{i1}}{\sum_k F^{k2} G^{k2}}\,, \quad \text{with}
\end{equation}
\begin{align}
& F^{ij}\,=\, |U_\nu^{ji}|^2
\quad \text{and} \ \
 G^{ij} \,=\, \left[m_P^2 (m_{\nu_i}^2+m_{l_j}^2) - 
 (m_{\nu_i}^2-m_{l_j}^2)^2 \right] \left[ (m_P^2 - m_{l_j}^2 -
  m_{\nu_i}^2)^2 - 4 m_{l_j}^2 m_{\nu_i}^2 \right]^{1/2}\,.
  \label{eq:FG}
\end{align}
The result of Eq.~(\ref{eq:RPresult}) has a straightforward
interpretation: $F^{ij}$ represents the impact of new interactions
(absent in the SM), whereas $G^{ij}$ encodes the mass-dependent
factors. The SM result can be easily recovered from
Eq.~(\ref{eq:RPresult}), in the limit $m_{\nu_i} = 0$ and $U_\nu^{ji}
= \delta_{ji}$,
\begin{equation} \label{eq:RMSM}
R_P^{SM} = \frac{m_e^2}{m_\mu^2}
\frac{(m_P^2-m_e^2)^2}{(m_P^2-m_\mu^2)^2} \,, 
\end{equation}
to which small electromagnetic corrections should be
added~\cite{Cirigliano:2007xi}.  

Using the results in Eqs.~\eqref{eq:RPresult} and \eqref{eq:RMSM}, we
obtain a general expression for $\Delta r_P$
\begin{equation}\label{eq:deltaRPresult}
\Delta r_P \,= \,\frac{m_\mu^2 (m_P^2 - m_\mu^2)^2}{m_e^2 (m_P^2 - m_e^2)^2}\,
\frac{\operatornamewithlimits{\sum}_{m=1}^{N_\text{max}^{(e)}} 
F^{m1}\, G^{m1}}
{\operatornamewithlimits{\sum}_{n=1}^{N_\text{max}^{(\mu)}} 
F^{m2}\, G^{n2}} -1 \,.
\end{equation}
Thus, depending on the masses of the new states (and their hierarchy)
and most importantly, on their mixings to the active neutrinos,
$\Delta r_P$ can considerably deviate from zero. In order to
illustrate this, we consider two regimes:

\begin{itemize}
\item {\bf Regime (A):} All sterile neutrinos are {\it lighter} than
  the decaying meson, but heavier than the active neutrino states,
  i.e. $m_\nu^\text{active} \ll m_{\nu_{s}} \lesssim m_P$
\item {\bf Regime (B):} All sterile neutrinos are {\it heavier} than
  $m_P$
\end{itemize}

Notice that in case (A), all the mass eigenstates can be kinematically
available and one should sum over all $3+N_s$ states; furthermore
there is an enhancement to $\Delta r_P$ arising from phase space
factors, see Eq. (\ref{eq:FG}).

\section{Constraints on sterile neutrinos}
\label{Sec:Constraints}

We review in this section the experimental and observational bounds on
the mass regimes and on the size of the active-sterile mixings that
must be satisfied.

First, it is clear that present data on neutrino masses and
mixings~\cite{Tortola:2012te} should be accounted for. Second, there
are robust laboratory bounds from direct sterile neutrinos searches
\cite{Atre:2009rg,Kusenko:2009up}, since the latter can be produced in
meson decays such as $\pi^\pm \to \mu^\pm \nu$, with rates dependent
on their mixing with the active neutrinos. Negative searches for
monochromatic lines in the muon spectrum can be translated into bounds
for $m_{\nu_s} - \theta_{i \alpha}$ combinations, where $\theta_{i
  \alpha}$ parametrizes the active-sterile mixing. The non-unitarity
of the leptonic mixing matrix is also subject to constraints. Bounds
on the non-unitarity parameter $\eta$ (Eq.~(\ref{eq:U:eta:PMNS})),
were derived using Non-Standard Interactions~\cite{Antusch:2008tz};
although not relevant in case (A), these bounds will be taken into
account when evaluating scenario (B).

The modified $W \ell \nu$ vertex also contributes to lepton flavour
violation (LFV) processes.  The radiative decay $\mu \to e \gamma$,
searched for by the MEG experiment~\cite{Adam:2011ch}, is typically
the most constraining observable~\footnote{Recently, it has been also
  noticed that in the framework of low-scale seesaw models, the
  expected future sensitivity of $\mu-e$ conversion experiments can
  also play a relevant r\^ole in detecting or constraining sterile
  neutrino
  scenarios~\cite{Ilakovac:2009jf,Dinh:2012bp,Alonso:2012ji,Ilakovac:2012sh}. This
  is also the case in the supersymmetric version of these models, even
  when the sterile neutrinos are heavier
  \cite{Hirsch:2012ax,Abada:2012cq}.}. The rate induced by sterile
neutrinos must satisfy~\cite{Ilakovac:1994kj,Deppisch:2004fa}
\begin{equation}\label{eq:BR:muegamma:sterile}
\text{BR}(\mu \to e \gamma) =  
\frac{\alpha_W^3 s_W^2 m_\mu^5}{256 \pi^2 m_W^4 \Gamma_\mu} |H_{\mu e}|^2
\leq 2.4 \times 10^{-12}\, ,
\end{equation}
where $H_{\mu e} = \sum_i U_\nu^{2i} U_\nu^{1i \, *} G_\gamma (
\frac{m_{\nu,i+3}^2}{m_W^2})$, with $G_\gamma$ the loop function and
$U_\nu$ the mixing matrix defined in Eq. \eqref{eq:cc-lag}.
Similarly, any change in the $W \ell \nu$ vertex will also affect
other leptonic meson decays, in particular $B \to \ell \nu$; the
following bounds were enforced in the analysis: $\text{BR}(B \to e
\nu) < 9.8 \times 10^{-7}$, $\text{BR}(B \to \mu \nu) < 10^{-6}$ and
$\text{BR}(B \to \tau \nu) = (1.65 \pm 0.34) \times
10^{-4}$~\cite{Beringer:1900zz}.

Important constraints can also be derived from LHC Higgs
searches~\cite{Dev:2012zg} and electroweak precision
data~\cite{delAguila:2008pw}. They will also be considered in our
numerical analysis.

Under the assumption of a standard cosmology, the most constraining
bounds on sterile neutrinos stem from a wide variety of cosmological
observations \cite{Smirnov:2006bu,Kusenko:2009up}. These include Large
Scale Structure data, X-ray searches (which can be produced in $\nu_i
\to \nu_j \gamma$), Lyman-$\alpha$ limits, the existence of additional
degrees of freedom at the epoch of Big Bang Nucleosynthesis and Cosmic
Microwave Background data. However, all the above cosmological bounds
can be evaded if a non-standard cosmology is considered. In fact, the
authors of Ref.~\cite{Gelmini:2008fq} showed that the above
cosmological constraints disappear in scenarios with low reheating
temperature. Therefore, we will allow for the violation of the latter
bounds, explicitly stating it.

\section{A numerical example: $\Delta r_K$ in the  inverse seesaw}
\label{Sec:drkiss}

Although the generic idea explored in this work applies to any model
where the active neutrinos have sizeable mixings with some additional
singlet states, we consider the case of the Inverse
Seesaw~\cite{Mohapatra:1986bd} to illustrate the potential of a model
with sterile neutrinos regarding tree-level contributions to light
meson decays.  As mentioned before, there are other
possibilities~\cite{Asaka:2005an,Ibarra:2010xw}.

\subsection{The inverse seesaw}

In the ISS, the SM particle content is extended by $n_R$ generations
of right-handed (RH) neutrinos $\nu_R$ and $n_X$ generations of
singlet fermions $X$ with lepton number $L=-1$ and $L=+1$,
respectively~\cite{Mohapatra:1986bd} (such that $n_R+n_X = N_s$).  In
our numerical application we will focus on the case $n_R = n_X = 3$.
The lagrangian is given by
\begin{equation}
\label{eq:L_IS}
\mathcal{L}_\text{ISS} = 
\mathcal{L}_{SM} + Y_{\nu}^{ij} \bar{\nu}_{R i} L_j \tilde{H}
+ {M_R}_{ij} \, \bar{\nu}_{R i} X_j + 
\frac{1}{2} {\mu_X}_{ij} \bar{X}^c_i X_j + \, \text{h.c.}
\end{equation}
where $i,j = 1,2,3$ are generation indices and $\tilde{H} = i \sigma_2
H^*$. Notice that the present lepton number assignment, together with
$L=+1$ for the SM lepton doublet, implies that the ``Dirac''-type
right-handed neutrino mass term $M_{R_{ij}}$ conserves lepton number,
while the ``Majorana'' mass term $\mu_{X_{ij}}$ violates it by two
units.

The left-handed neutrinos mix with the right-handed ones after
electroweak symmetry breaking. This leads to an effective Majorana
mass for the active (light) neutrinos. Assuming $\mu_X \ll m_D \ll
M_R$, where $m_D= \frac{1}{\sqrt 2} Y_\nu v$, with $v$ the vacuum
expectation value of the SM Higgs boson, one obtains
\begin{equation}\label{eq:nu}
m_\nu \simeq {m_D^T M_R^{T}}^{-1} \mu_X M_R^{-1} m_D \, .
\end{equation}
The remaining 6 sterile states have masses approximately given by
$M_{\nu} \simeq M_R$. Small corrections can be added to these results,
but they are typically negligible \cite{Forero:2011pc}.

In what follows, and without loss of generality, we work in a basis
where $M_R$ is a diagonal matrix (as are the charged lepton Yukawa
couplings). $Y_\nu$ can be written using a modified Casas-Ibarra
parametrisation~\cite{Casas:2001sr} (thus automatically complying with
light neutrino data),
\begin{equation}\label{eq:CI:param}
Y_\nu = \frac{\sqrt{2}}{v} \, V^\dagger \, \sqrt{\hat M} \, R \,
\sqrt{{\hat m}_\nu} \, U_\text{PMNS}^\dagger \, ,
\end{equation}
where $\sqrt{{\hat m}_\nu}$ is a diagonal matrix containing the square
roots of the three eigenvalues of $m_\nu$ (cf. Eq. \eqref{eq:nu});
likewise $\sqrt{\hat M}$ is a (diagonal) matrix with the square roots
of the eigenvalues of $M = M_R \mu_X^{-1} M_R^T$.  $V$ diagonalizes
$M$ as $V M V^T = \hat{M}$, and $R$ is a $3 \times 3$ complex
orthogonal matrix, parametrized by $3$ complex angles, encoding the
remaining degrees of freedom.

The nine neutrino mass eigenstates enter the leptonic charged current
through their left-handed component (see Eq. \eqref{eq:cc-lag}, with
$i = 1, \dots, 9$, $j = 1, \dots, 3$). The unitary leptonic mixing
matrix $U_\nu$ is now defined as $U^T_\nu \mathcal{M} U_\nu =
\text{diag}(m_i)$. Notice however that only the rectangular $3 \times
9$ sub-matrix (first three columns of $U_\nu$) appears in
Eq. \eqref{eq:cc-lag} due to the gauge-singlet nature of $\nu_R$ and
$X$.

\subsection{Numerical evaluation of $\Delta r_K$ in the inverse seesaw}

We numerically evaluate the contributions to $R_K$ in the framework of
the ISS and address the two scenarios discussed before, which can be
translated in terms of ranges for the (random) entries of the $M_R$
matrix: {\it regime (A)} ($m_{\nu_s} < m_P$) - ${M_R}_{i} \in
[0.1,200]$ MeV; {\it regime (B)} ($m_{\nu_s} > m_P$) - ${M_R}_{i} \in
[1,10^6]$ GeV.  The entries of $\mu_X$ have also been randomly varied
in the $[0.01$ eV$, 1$ MeV$]$ range for both cases.

The adapted Casas-Ibarra parametrisation for $Y_\nu$,
Eq.~(\ref{eq:CI:param}), ensures that neutrino oscillation data is
satisfied (we use the best-fit values of the global analysis
of Ref.~\cite{Tortola:2012te} and set the CP violating phases of
$U_\text{PMNS}$ to zero). The $R$ matrix angles are taken to be real
(thus no contributions to lepton electric dipole moments are
expected), and randomly varied in the range ${\theta}_{i} \in [0,2
  \pi]$. We have verified that similar $\Delta r_K$ contributions are
found when considering the more general complex $R$ matrix case.

\begin{figure}[ht]
\begin{tabular}{cc}
\hspace*{5mm}{\footnotesize Scenario (A)} &
\hspace*{11mm}{\footnotesize Scenario (B)}\vspace*{2mm} \\
\includegraphics[width=75mm]{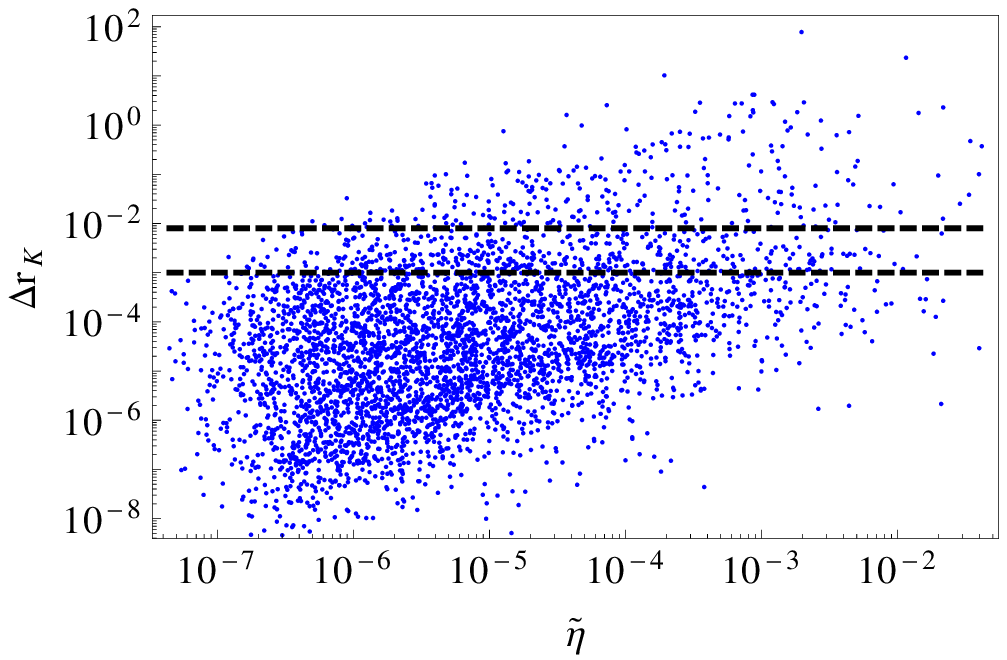}
\hspace*{2mm} 
&
\hspace*{2mm}
\includegraphics[width=75mm]{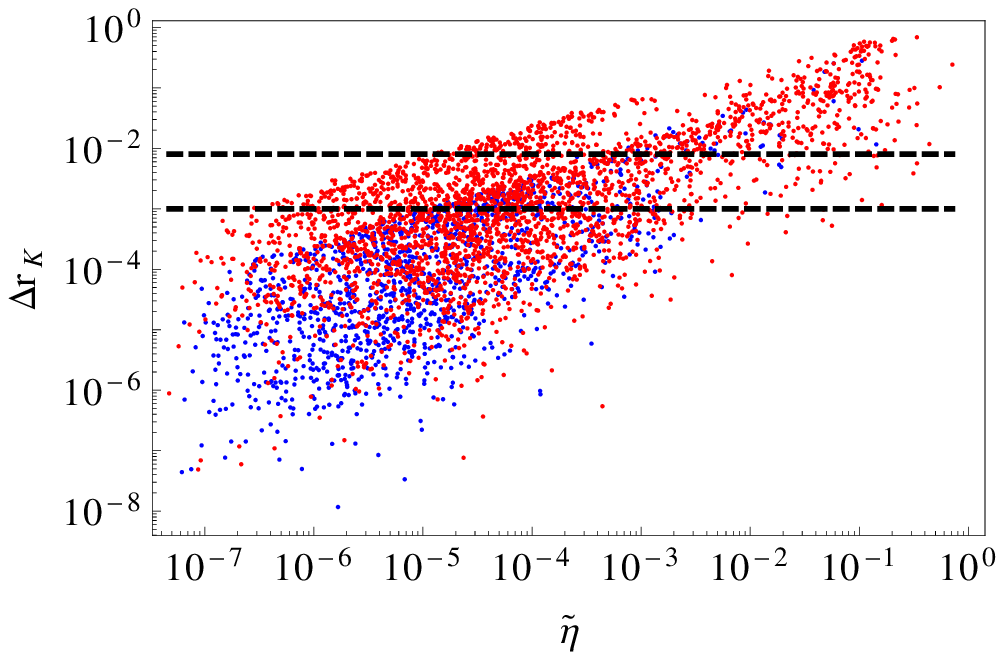} 
\end{tabular}
\caption{Contributions to $\Delta r_K$ in the inverse seesaw as a
  function of $\tilde \eta = 1 - |\text{Det}(\tilde U_\text{PMNS})|$:
  regimes A (left) and B (right). The upper (lower) dashed line
  denotes the current experimental limit (expected sensitivity).  On
  the right panel, red points denote cases where $Y_\nu \gtrsim
  10^{-2}$.  All points comply with experimental and laboratory
  constraints.  Points in (B) are also in agreement with cosmological
  bounds, while those in (A) require considering a non-standard
  cosmology.  }
\label{figure12}
\end{figure}

In Figs.~\ref{figure12}, we collect our results for $\Delta r_K$ in
scenarios (A) - left panel - and (B) - right panel, as a function of
$\tilde \eta$, which parametrizes the departure from unitarity of the
active neutrino mixing sub-matrix $\tilde U_\text{PMNS}$, $\tilde \eta
= 1 - |\text{Det}(\tilde U_\text{PMNS})|$.  Although the cosmological
constraints are not always satisfied, we stress that all points
displayed comply with the different experimental and laboratory bounds
discussed before.
For the case of scenario (A), one can have very large contributions to
$R_K$, which can even reach values $\Delta r_K \sim \mathcal{O}(1)$
(in some extreme cases we find $\Delta r_K$ as large as $\sim 100$).
The hierarchy of the sterile neutrino spectrum in case (A) is such
that one can indeed have a significant amount of LFU violation, while
still avoiding non-unitarity bounds.  Although this scenario would in
principle allow to produce sterile neutrinos in light meson decays,
the smallness of the associated $Y_\nu$
($\lesssim\mathcal{O}(10^{-4})$), together with the loop function
suppression ($G_\gamma$), precludes the observation of LFV processes,
even those with very good associated experimental sensitivity, as is
the case of $\mu \to e \gamma$.  The strong constraints from CMB and
X-rays would exclude scenario (A); in order to render it viable, one
would require a non-standard cosmology.

Despite the fact that in case (B) the hierarchy of the sterile states
is such that non-unitarity bounds become very stringent (since the
sterile neutrinos are not kinematically viable meson decay final
states), sizeable LFU violation is also possible, with deviations from
the SM predictions again as large as $\Delta r_K \sim
\mathcal{O}(1)$. Although one cannot produce sterile states in meson
decays in this case, the large $Y_\nu$ open the possibility of having
larger contributions to LFV observables so that, for example, BR($\mu
\to e \gamma)$ can be within MEG reach.

Although we do not explicitly display it here, the prospects for
$\Delta r_\pi$ are similar: in the same framework, one could have
$\Delta r_\pi \sim \mathcal{O}(\Delta r_K)$, and thus $\Delta r_\pi
\sim \mathcal{O}(1)$ in both scenarios.  Depending on the singlet
spectrum, these observables can also be strongly correlated: if all
the sterile states are either lighter than the pion (as it is the case
of scenario (A)) or then heavier than the kaon, one finds $\Delta
r_\pi \approx \Delta r_K$. This is a distinctive feature of our
mechanism.

\section{Concluding remarks}
\label{Sec:conclusions}

The existence of sterile neutrinos can potentially lead to a
significant violation of lepton flavour universality at tree-level in
light meson decays. As shown in this study, provided that the
active-sterile mixings are sufficiently large, the modified $W \ell
\nu$ interaction can lead to large contributions to lepton flavour
universality observables, with measurable deviations from the standard
model expectations, well within experimental sensitivity.  This
mechanism might take place in many different frameworks, the exact
contributions for a given observable being model-dependent.

As an illustrative (numerical) example, we have evaluated the
contributions to $R_K$ in the inverse seesaw extension of the SM - a
truly minimal extension of the SM - , for distinct hierarchies of the
sterile states. Our analysis reveals that very large deviations from
the SM predictions can be found ($\Delta r_K \sim \mathcal{O}(1))$ -
or even larger, well within reach of the NA62 experiment at CERN.
This is in clear contrast with other models of new physics (for
example unconstrained SUSY models, where one typically has $\Delta r_K
\lesssim \mathcal{O}(10^{-3})$).

\ack

The author is grateful to A. Abada, D. Das, A. M. Teixeira and
C. Weiland for their collaboration in the work this talk is based on.
This work has been supported by the ANR project CPV-LFV-LHC
NT09-508531.

\end{document}